# Hourly Traffic Prediction of News Stories

Luís Marujo
LTI/CMU, USA
INESC-ID/IST, Portugal
Luis.Marujo@inesc-id.pt

Miguel Bugalho
INESC-ID/IST, Portugal
Miguel.Bugalho@l2f.inesc-id.pt

João P. Neto
INESC-ID/IST, Portugal
Joao.Neto@inesc-id.pt

Anatole Gershman
LTI/CMU, USA
anatoleg@cs.cmu.edu

Jaime Carbonell
LTI/CMU, USA
jgc@cs.cmu.edu

## ABSTRACT
The process of predicting news stories popularity from several news sources has become a challenge of great importance for both news producers and readers. In this paper, we investigate methods for automatically predicting the number of clicks on a news story during one hour. Our approach is a combination of additive regression and bagging applied over a M5P regression tree using a logarithmic scale ($\log_{10}$). The features included are social-based (social network metadata from Facebook), content-based (automatically extracted keyphrases, and stylometric statistics from news titles), and time-based. In 1st Sapo Data Challenge we obtained 11.99% as mean relative error value which put us in the 4th place out of 26 participants.

## Categories and Subject Descriptors
H.3.3 [**Information Search and Retrieval**]: Information filtering

## 1. INTRODUCTION
"Can we predict the number of clicks that a news story link receives during one hour"? This was the main research question proposed in the *1st Sapo Data Challenge – Traffic prediction*[1] launched by PT Comunicações. Sapo is the largest Portuguese web portal. Its home page (http://www.sapo.pt) receives about 13 million daily page views and 2.5 million daily visits[2]. The home page has several sections that link to different types of content, such as news, videos, opinion articles, blog previews, etc.

Currently, the selection and ordering of news stories is done mostly manually by the site editors. Obviously, manual solutions do not scale and we need to find a method for automatically predicting popularity of news stories.

[1] http://labs.sapo.pt/blog/2011/03/10/1st-sapo-data-challenge/

[2] Based on Alexa estimates, as of Jul 7th 2011

Predicting the popularity of a news story is a difficult task [9]. Popularity of a story can be measured in terms of the number of views, votes or clicks it receives in a period of time. Click-through rate (CTR) is the most popular way of measuring success [7, 10]. It is defined as the ratio of the number of times the user clicked on a page link and the total number of times the link was presented.

Popularity of a news story is influenced by many factors, including the item's quality, social influence and novelty. The item's quality is mixture of fluency, rhetoric devices, vocabulary usage, readability level, and the ideas expressed which makes quality hard to measure [15]. The social influence consists on knowing about other people's choices and opinions [9]. Salganik et al. [15] show that item's quality is a weaker predictor of popularity than social influence. This partially explains the difficulty of predicting article popularity based solely on its content and novelty.

Most popular portals such as Digg and Slashdot allow users to submit and rate news stories by voting on them. This information is often used by a collaborative filtering algorithm to predict popularity, select and order news items [8, 9, 16]. Typically, these models perform linear regression [18] on a logarithmically transformed data.

In this paper, we present an approach for predicting the click rate based on a combination of content-based, social network-based, and time-based features. The main novelty of the approach is the type of content-based features we extract (e.g.: number of keyphrases), the inclusion of time-based features, and the prediction process that combines several regression methods to produce and estimate of a number of clicks per hour.

This paper is organized as follows: Section 2 presents the dataset used to train and test our predictions system; the description of the proposed prediction methodology is presented in Section 3; the results are described in Section 4, and Section 5 contains conclusions and suggestions for future work.

## 2. Dataset
The dataset contains 13140 *link-hour* entries from 1217 news story links gathered from several news sources over 15 consecutive days. Each link-hour entry records the number of clicks on a particular link shown in Sapo Portal (Figure 1) during a particular hour. Each entry contains 8 fields:

1. **Line Number**: number identifying the entry.

2. **Date + Time information**: the date and daytime at which the hits took place as a string.
3. **Channel ID**: a number identifying Sapo's source (channel) that produced the content. There are contents from 18 different sources.
4. **Section (topic)**: there are 5 possible sections: general, sport, economy, technology, and life.
5. **Subsection**: each placeholder is further divided in five subsections: "manchete", "headlines" and "related", "footer" and "null". This is an important parameter because each subsection is visually smaller than the previous ones when presented to the user (Figure 1).
6. **News ID**: an integer identifying the content.
7. **Number of hits/clicks**: the number of hits that the linked content received, during one hour (see field 2 above).
8. **Title**: the title of the news story.

Example: *[13116] [2011-03-08 23:00:00] [2] [geral] [manchete] [1214] [401] [Barcelona segue para os quartos-de-final]*

At first, there was a training set (95% of all entries) and then a test set became available.

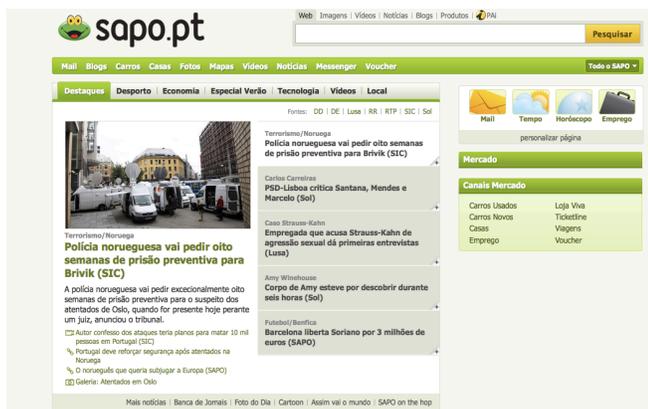

**Figure 1: Sapo Portal (http://www.sapo.pt).**

Figure 2 shows the long tail distribution [1] of number of clicks per entry. Figure 3 conveys the same information in a logarithmic scale. We included Figure 4 to show the distribution of the number of clicks per hour, which influences our prediction methods. Figure 5 shows the average number of clicks per hour. It is interesting to note that 9 a.m., 12 a.m., and 10 p.m. are the peak hours. During day hours the average number of clicks is more or less constant, i.e. the boundary between working hours and leisure hours is not visible.

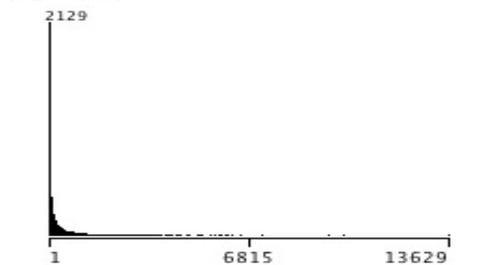

**Figure 2: Distribution of Number of Clicks (Y-axis contains number of entries and X-axis is the number of clicks)**

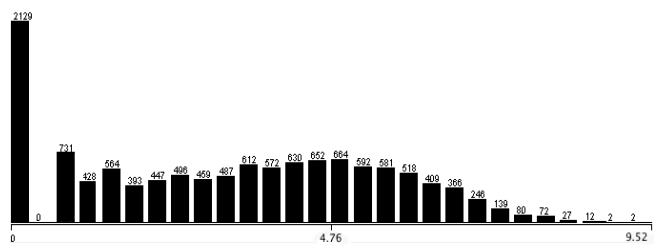

**Figure 3: Natural Logarithmic Distribution of Number of Clicks.**

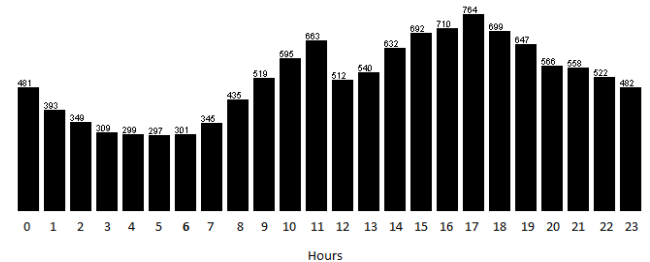

**Figure 4: Distribution of link-hour entries per hour over 15 days.**

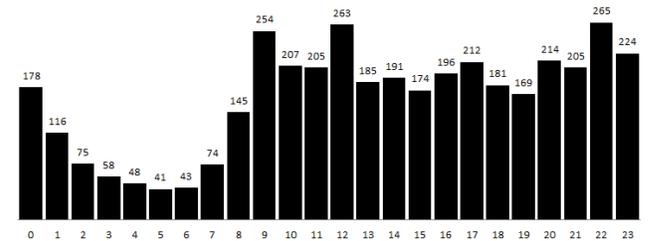

**Figure 5: Average number of clicks per hour.**

## 3. Prediction Methodology

To address the prediction of clicks challenge, we adopted a supervised learning framework, based on WEKA [4],. It consists of 2 steps: feature extraction, and regression.

### 3.1 Feature Extraction

Figure 6 provides an overview of the feature extraction process which starts with the initial (*base*) features taken from the dataset entries and produces an enriched set of 3 types of features: content-based, social network-based, and time-based. The content-based features include: the number of web pages containing the same title (*F1*), the number of occurrences of certain key phrases in news articles (*F3*), and the stylometric features of the title (*F4*).

**Content-based features (F1, F3, and F4)**

We use the news title as query in the Sapo RSS search interface to get the total number of stories with the same title (*F1*).

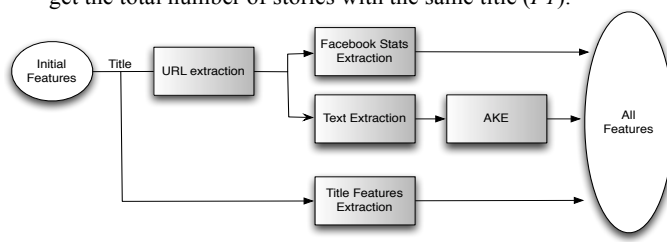

**Figure 6: Feature Extraction Enrichment Process.**

The main textual content is extracted using the boilerpipe library [5]. Our supervised Automatic Key-phrase Extraction (AKE) [11] method is applied over the extracted news text from the training set documents to create a list key phrases. The AKE system, developed for European Portuguese Broadcast News [11], is an extended version of Maui-indexer toolkit [12], which in turn is an improved version of KEA [19]. The system can be easily adapted to support other languages such as English.

We filtered key phrases, extracted using the AKE system, with a confidence level lower than 50%. Their number of occurrences is used as features of an article (*F3*). The final list of key phrases contains 34 key phrases, *e.g.: Portugal, United States, market.*

These key phrases are used to compare the content of news stories.

The stylometric features of the title (*F4*) include: the number of words, maximum word length, minimum word length, the number of quotes, the number of capital letters, and the number of named entities identified by MorphAdorner Name Recognizer[3].

**Social network-based features (F2)**

The social network-based features are metadata information retrieved from Facebook (*F2*). The social-based features are extracted by calling the Facebook API and retrieving Facebook Metadata or statistics containing the URL of the article: the number of shares, the number of likes, the number of comments, and the total number of occurrences in Facebook data. We have also extracted Twitter metadata, i.e., the number of tweets, but it was excluded because Portuguese news tweets containing URLs are rare.

**Time-based features (F5)**

There are 3 time-based features: day, time and the number of hours elapsed from the initial publication of the article (*F5*).

The number of hours elapsed from the initial publication of an article (*F5*) is used as the initial approximation of its novelty.

## 3.2 Regression

The goal of regression methods is to build a function *f(x)* that maps a set of independent variables or features (*X1, X2,..., Xn*) into a dependent variable or label *Y*. In our case, we aim to build regression models using a training dataset to predict the number of clicks in the test set.

In this work we explored a combination of regression algorithms. We explored linear regression, regression trees: REPTree and M5P. To further improve the results, we combined the best performing regression algorithm (M5P) with two meta-algorithms: Additive Regression and Bagging.

### REPTree – regression-based tree
The REPTree algorithm is fast regression tree learner based on C4.5 [18]. It builds a regression tree using information variance, reduced-error pruning (with back-fitting), and only sorting numeric attributes once.

### M5P – regression-based tree
The M5P algorithm is used for building regression-based trees[14][17]. M5P is a reconstruction of Quinlan's M5 algorithm for inducing trees of regression models. M5P combines a conventional decision tree with the possibility of linear regression functions at the nodes. First, a decision-tree induction algorithm is used to build a tree, but then, instead of maximizing the information gain at each inner node, a splitting criterion is used that minimizes the intra-subset variation in the class values down each branch. The splitting procedure in M5P stops if the class values of all instances that reach a node vary very slightly, or only a few instances remain.

### Bagging
Bagging, also known as Bootstrap aggregating, is a machine learning meta-algorithm proposed by Breiman [2] which is used with many classification and regression techniques, such as decision tree models to reduce the variance associated with the predictions, thereby improving the results.

The idea consists of using multiple versions of a training set; each version is created by randomly selecting samples of the training dataset, with replacement. For each subset, a regression model is built by applying a previously selected learning algorithm. The learning algorithm must remain the same in all iterations. The final prediction is given by averaging the predictions of all the individual classifiers.

### Additive Regression (AR)
Additive regression, or Stochastic Gradient boosting [3], is a machine learning ensemble meta-algorithm or meta-classifier. It enhances the performance of another regression classifier (base or weak learner) $h_i(x)$ such as regression tree.

This method is an iterative algorithm, which constructs additive models (sum of weak learners – Equation 2) by fitting a base learner to the current residue at each iteration. The residue is the gradient of the loss function *L(t_i, s_i)*; where *t_i* is the true value of the sample and *s_i* = *H*_{k-1}(*x_i*). The default loss function is least squares. At each iteration, a subsample of data is drawn uniformly at random, without replacement, from the full training set. This random subsample is used to train the base learner $h_i(x)$ to produce a model (Equation 2) for the current iteration.

$$H_k(x) = \sum_{i=0}^{k} \beta_i \, h_i(x) \qquad (2)$$

where $\beta_i$ denotes the learning rate (expansion coefficients), k is the number of iterations and *x* is the features vector.

### Combining Bagging and Additive Regression
Because bagging is more robust than additive regression in noisy datasets, but additive regression performs better in reduced or noise-free data, Sotiris [6] proposed the combination of bagging and additive regression and showed improvements for regression trees. Our combination approach consists of using the bagging meta-classifiers as the base classifiers of the additive regression.

## 4. Experimental Evaluation
In this section, we describe the evaluation procedures used during this work. We divided the evaluation in 2 stages: at first, the test set was not available and we used the training set for the experiments using 10-fold cross-validation. When the test set became available, we trained on the whole training set and evaluated on the test set. We evaluated our results using *Mean Absolute Error* and *Relative Absolute Error*. Assume that $p_i$ is the predicted number of clicks and $t_i$ is the true value, then:

---

[3] http://morphadorner.northwestern.edu

$$Absolute\ Error: AE(i) = |p_i - t_i| \qquad (3)$$

$$Relative\ Error: RE(i) = \frac{AE(i)}{t_i} \qquad (4)$$

$$Cumulative\ Absolute\ Error: CAE = \sum_{i=1}^{n} AE(i) \qquad (5)$$

$$Cumulative\ Relative\ Error: CRE = \sum_{i=1}^{n} RE(i) \qquad (6)$$

$$Mean\ Absolute\ Error: MAE = \frac{CAE}{n} \qquad (7)$$

$$Mean\ Relative\ Error: MRE = \frac{CRE}{n} \qquad (8)$$

The challenges results were presented in Cumulative Absolute Error and Cumulative Relative error. In this paper we opted to provide their means to make the results comparable (cumulative results depend on the number of examples). It is important to note that relative absolute error displayed in the WEKA interface differs from the equations above, and as a result it was not used in this work. In the WEKA interface this measure divides the mean absolute error by the corresponding error of the ZeroR classifier on the data (i.e.: the classifier predicting the prior probabilities of the classes or values observed in the data).

Table 1 shows the results using several supervised machine-learning techniques obtained in the first evaluation stage. This evaluation was performed in the training set using 10 fold cross validation. We tried both REPTree and linear regression, because they have a faster training time. In addition, linear regression is the most frequent method selected for popularity estimation. Nevertheless, they were outperformed by M5P.

M5P regression generates outlier predictions, i.e.: a negative number of clicks or very large number of clicks. We have considered two possibilities to solve this problem: either setting all outliers to 1; or changing the negative outliers values to 1 and the positive outliers to the maximum number of clicks. The first solution gave better results. The lack of information justifies the occurrence of these outliers. Therefore, we used the most frequent value (1) given the number of clicks distribution (Figure 2).

The conversion of the number of clicks to a logarithmic scale helped to approximate a linear distribution to which our methods could be better suited. This fact is easy to understand by comparing the distribution of clicks in Figure 2 and 3. In addition, it also helped to eliminate outliers.

**Table 1 – Results obtained in the training set using all features and 10 fold cross-validation (p-value ≈ 0.0).**

| Configuration | MAE | MRE |
|---|---|---|
| Linear Regression | 223.09 | 49.07% |
| REPTree | 199.09 | 27.31% |
| M5P | 189.86 | 27.51% |
| AR+Bag.+M5P | 183.14 | 26.84% |
| AR+Bag.+M5P + (outliers values set to 1) | 171.30 | 24.65% |
| AR+Bag.+M5P+$log_{10}$ | 158.85 | 5.04 % |
| AR+Bag.+M5P+ln | **158.46** | **5.00%** |

Table 2 captures the influence of the enrichment process by showing the improvements of adding new features. The values were obtained using M5P on the training set and 10 fold cross-validation. The improvements obtained using social media features (F2) were limited. This means a low correlation between the sharing of Portuguese news in Social Networks and the number of accesses to Sapo portal.

The best performing features were the keyphrase based features. They capture semantic information at a more detailed level than topic information does. Information about locations, such as Lisbon and Oporto (the capital of Portugal and the second most important city in Portugal); sports clubs e.g.: Benfica, (Futebol Clube do) Porto, politics e.g.: European Union, Greece, Europe; economics, e.g.: fee, market; and technology, e.g.: computers.

The stylometric features extracted from the title (*F4*) were also useful to reduce both MAE and MRE. The number of hours from the initial publication (*F5*) conveys a very small improvement in the results. We observed improvement with the Base features and when we include bagging and additive regression.

**Table 2 - Results in the training set using M5P and 10 fold cross-validation (p-value ≈ 0.0).**

| Features | MAE | MRE |
|---|---|---|
| Base | 223.92 | 49.68% |
| Base+F1 | 222.78 | 48.81% |
| Base+F1+F2 | 222.37 | 48.88% |
| Base+F1+F2+F3 | 194.10 | 29.30% |
| Base+F1+F2+F3+F4 | **189.86** | **27.51%** |
| All | **189.86** | **27.51%** |

Table 3 present the results obtained in the test set in MAE and MRE, but the challenge used the cumulative absolute error and cumulative relative error. The results for all participants are available in the challenge final results webpage[4]. Our prediction system obtained the 4th place out of 26 participants from 9 countries.

Without applying the logarithmic transformation, our system falls into 7th place.

We noticed the MRE for the test set was higher. Close examination of the test set revealed that some of its characteristics were different from the training set. For example, the average number of clicks and the variance in the number of clicks were significantly higher in the training set which caused the algorithm overestimate the number of clicks in its predictions for the test set. That may explain the discrepancy in the results.

**Table 3 - Results in the test set (submitted to the Challenge) (p-value ≈ 0.0)**

| Configuration | MAE | MRE |
|---|---|---|
| AR+Bag.+M5P | 226.84 | 54.54% |
| AR+Bag.+M5P+ln | 152.86 | 12.38% |
| AR+Bag.+M5P+$log_{10}$ | **152.59** | **11.99%** |

---

[4] http://labs.sapo.pt/blog/2011/06/20/1st-sapo-data-challenge-final-results/

## 5. Conclusions

In this paper we described the problem of predicting the number of clicks that a news story link receives during one hour. Real data from the largest Portuguese web portal, Sapo, was used to train and test our proposed prediction methodology. The data was made available to all participants of the 1st Sapo Data Challenge - Traffic Prediction.

Despite the fact that predicting item's popularity per hour is a very difficult task, our approach obtained results that are close to the real number of clicks (12% MRE and 152 MAE). These results yield us a 4th place (in all categories) in the challenge over 26 participants.

Results have shown that the social-based and time-based features had little correlation with the number of clicks in the portal. In contrast, the content-based features had a very large impact. This contradicts the results obtained by Lerman [8,9], which, perhaps, can be explained by their use of Digg, a social media news site.

Regarding the method, the usage of logarithmic scale on the number of clicks had the greatest impact on the final result, especially on MRE (almost 20% improvement on the training set using 10 fold cross validation). However, both the refinement of the regression methods used and the constant setting of the outliers obtained visible improvements. In fact for the MAE results, improvements of 16 and 12 percentage points were obtained for the regression refinement and outlier treatment, which, in total, are more than the double of the improvement obtained with the logarithmic scale transformation (that yielded a gain of 13 percentage points).

In future work, we will investigate ways to increase the use of automatic key-phrase extraction, e.g.: including a larger set of concepts that better capture the document contents and topics. We will also explore if the inclusion of sentiment analysis features can improve the accuracy of our predictions.

## 6. ACKNOWLEDGMENTS

The authors would like to thank PT Comunicações for providing the dataset and creating the 1st Sapo Data Challenge – Traffic Prediction. We want to thank Professors Mário Figueiredo and Isabel Trancoso for fruitful comments. Support for this research by FCT through the Carnegie Mellon Portugal Program and under FCT grant SFRH/BD /33769/2009. This work was also partially funded by European Commission under the contract FP7-SME-262428 EuTV, QREN SI IDT 2525, and SI IDT 5108. Support by FCT (INESC-ID multiannual funding) through the PIDDAC Program funds.